\documentclass[
    aps,
    prl,
    reprint,
    superscriptaddress,
    amsmath,
    amssymb,
    floatfix,
]{revtex4-2}

\usepackage{graphicx}
\usepackage{hyperref}
\usepackage{xurl}
\hypersetup{hidelinks}

\begin{document}

\title{Physical Fidelity of Wave-Based Inverse Problems under \\Broken Time-Reversal Symmetry}

\author{Haiyan Ou}
\email{ouhaiyan@uestc.edu.cn}
\affiliation{School of Physics, University of Electronic Science and Technology of China, Chengdu 611731, China}
\affiliation{Shenzhen Institute for Advanced Study, University of Electronic Science and Technology of China, Shenzhen, China}

\author{Jiajia Duan}
\affiliation{Shenzhen Institute for Advanced Study, University of Electronic Science and Technology of China, Shenzhen, China}

\author{Edmund Y. Lam}
\affiliation{Department of Electrical and Computer Engineering, The University of Hong Kong, Pokfulam, Hong Kong, China}

\author{Bingzhong Wang}
\email{bzwang@uestc.edu.cn}
\affiliation{School of Physics, University of Electronic Science and Technology of China, Chengdu 611731, China}

\date{September 22, 2026}

\begin{abstract}

Modern coherent imaging commonly treats defocus interference as an undesirable artifact to be suppressed. Here, we show that this interference is instead a deterministic projection shadow associated with broken time-reversal symmetry (TRS) under finite-aperture observation. The resulting non-unitary observation operator imposes irretrievable information loss, making this shadow an unavoidable coherent residual. We demonstrate that suppressing this residual beyond the measurement support can yield physically inconsistent reconstructions by assigning unconstrained components within the null space, thereby disrupting the propagation of the reconstructed complex field. We term this failure mode physical overfitting and introduce the complex-field TRS residual metric, \(R_{\mathrm{TRS}}(z)\). By using the measured axial evolution of the projection shadow as a \(z\)-resolved physical reference, the metric extends fidelity assessment from two-dimensional (2D) detector-plane agreement to propagation-resolved three-dimensional (3D) complex-field consistency. These results establish a physical constraint on phase-space retrieval and wave-based inverse reconstruction.

\end{abstract}

\maketitle

\section{Introduction}

\begin{figure}[b]
    \centering
    \includegraphics[width=\columnwidth]{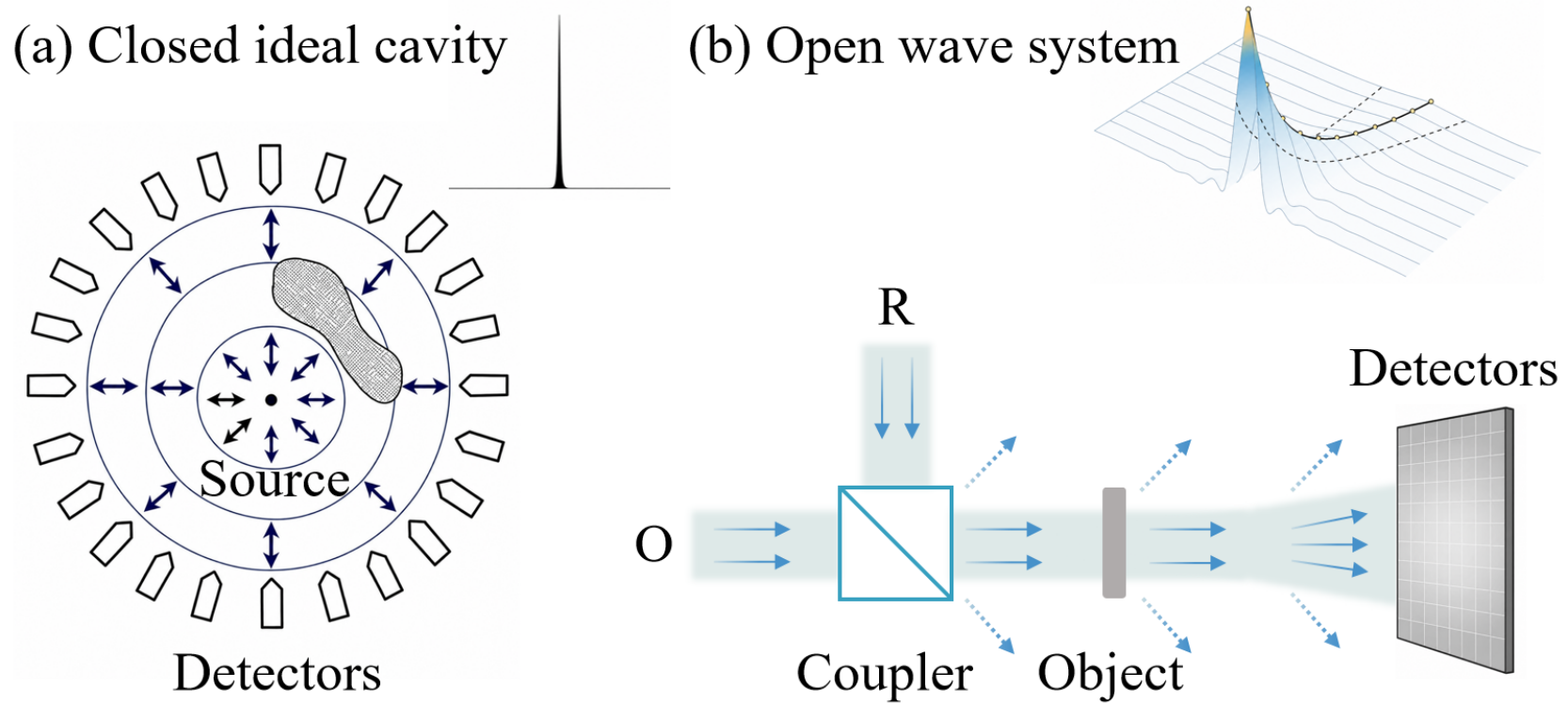}

    \caption{\textbf{Time reversal in closed versus open wave systems.}
    (a) An ideal cavity provides omnidirectional observation, preserving TRS to yield a diffraction-limited focus (inset). (b) Finite angular acceptance in an open system introduces a spatial truncation operator $\mathcal{W}$ that breaks TRS. The schematic illustrates a coherent detection setup, where O and R denote the object and reference waves, respectively. This restricted observation process limits the adjoint reconstruction, producing an extended 3D point-spread function and its deterministic projection shadow (inset).}
    \label{fig:fig1}
\end{figure}

In an ideal closed system with complete field access, time reversal achieves physically optimal reconstruction~\cite{Fink1997Time}.
The backward propagation of a wave field is governed by an adjoint operator, allowing the restoration of the initial source state as a diffraction-limited focus without extended macroscopic interference~[Fig.~\ref{fig:fig1}(a)].
However, practical coherent imaging systems are effectively open with respect to the accessible field modes due to finite observation apertures [Fig.~\ref{fig:fig1}(b)]. This spatial truncation renders the
effective observation operator non-unitary and leads to the irreversible loss of high-spatial-frequency components and, where applicable, evanescent content. This loss inevitably manifests as an extended 3D point spread function (PSF). For decades, this diffractive broadening has been acknowledged as the physical limit of spatial resolution~\cite{Born_principle_of_optics_1999}. Consequently, the resulting background distribution, commonly known as defocus noise, blur, or interference, is conventionally treated as stochastic error or undesirable artifact targeted for numerical suppression~\cite{bertero1998introduction}.

Driven by this perspective, many reconstruction frameworks, including sparsity- or smoothness-based regularization~\cite{RUDIN:1992,Candes:06} and deep learning~\cite{Barbastathis:19,Zeng:21}, achieve remarkable visual clarity by mapping degraded measurements to idealized 2D distributions. These image-centric techniques optimize statistical properties to eliminate background interference. However, this approach overlooks the deterministic link between spatial truncation and the resulting complex field. When visual fidelity is optimized at the expense of the physically supported coherent residual, such methods can compromise the physical consistency of reconstructed wave evolution.

This work redefines defocus interference as a projection shadow of broken time-reversal symmetry (TRS) induced by spatial truncation in open systems, rather than a breakdown of the underlying time-reversal-invariant wave dynamics. By introducing the TRS residual metric, \(R_{\mathrm{TRS}}(z)\), a criterion for physical fidelity is established beyond 2D detector-plane agreement. The evolution of this shadow is utilized as a reference for testing the consistency of reconstructed complex fields with wave propagation. Simulations and experiments show that its algorithmic suppression can induce physical overfitting by introducing unconstrained null-space components. The resulting reconstructions can appear favorable in 2D while remaining inconsistent with propagation-resolved complex-field evolution.

\section{Finite-Aperture Time Reversal and the Projection Shadow}

Consider a scalar monochromatic source \(s(\omega)\) at \(\mathbf r_0\) in a lossless reciprocal medium. On the closed surface \(\Gamma\), the forward operator \(\mathcal H\), with kernel \(G\), yields \(u_\Gamma(\mathbf r,\omega)=s(\omega)G(\mathbf r,\mathbf r_0;\omega)\). Under the \(e^{-i\omega t}\) convention, ideal time reversal phase-conjugates and re-emits the complete boundary field, \(u_\Gamma^*(\mathbf r,\omega)=s^*(\omega)G^*(\mathbf r,\mathbf r_0;\omega)\).

Applying the adjoint operator $\mathcal{H}^\dagger$, these conjugated waves backward propagate into the cavity. The reconstructed field $u_{\text{TR}}$ at an arbitrary point $\mathbf{r}'$ is formulated as the superposition integral:

\begin{equation}
u_{\text{TR}}(\mathbf{r}', \omega) = s(\omega) \int_{\Gamma} G(\mathbf{r}', \mathbf{r}; \omega) G^*(\mathbf{r}, \mathbf{r}_0; \omega) dS
\label{eq:ideal_TR}
\end{equation}
This integral defines the complete-aperture time-reversal correlation kernel. At a fixed frequency, it produces a diffraction-limited focus centered at $r_0$, rather than an exact spatial Dirac delta function~\cite{Carminati:07,Fink:92}. In the ideal complete-mode limit, the forward--adjoint map satisfies $\mathcal H^\dagger\mathcal H=\mathcal I$; time reversal is then information-preserving within the accessible wave space.

The complete-aperture correlation between fields generated at two positions $\mathbf{r}_1$ and $\mathbf{r}_2$ is:

\begin{equation}
\begin{aligned}
\mathcal{C}_\Gamma(\mathbf{r}_1, \mathbf{r}_2)
&\equiv \langle \mathcal{H}(\mathbf{r}_1), \mathcal{H}(\mathbf{r}_2) \rangle_\Gamma \\
&= \int_\Gamma G^*(\mathbf{r}, \mathbf{r}_1; \omega)
G(\mathbf{r}, \mathbf{r}_2; \omega)\,dS
\end{aligned}
\label{eq:ideal_orthogonal}
\end{equation}
At a fixed frequency, \(\mathcal C_{\Gamma}\) is generally nonzero for \(\mathbf r_1\neq\mathbf r_2\) and defines the diffraction-limited reference correlation of complete-aperture observation.

Practical coherent imaging setups, however, access only a finite portion of \(\Gamma\). Let \(W\) denote the corresponding aperture window and \(\mathcal H_A=W\mathcal H\) the effective observation operator. For a binary aperture, \(W=W^\dagger=W^2\) and \(W\neq I\)~\cite{Fink:93}. The adjoint reconstruction of an arbitrary  recorded field \(f\) is:

\begin{equation}
    \hat f=\mathcal H_A^\dagger\mathcal H_Af=\mathcal H^\dagger  \mathcal{W}\mathcal H f
\end{equation}
resulting in an imperfect reconstruction $\hat{f} \neq f$. Even in a lossless medium, finite-aperture observation removes inaccessible modes and makes the effective reconstruction operator non-unitary and non-invertible on its null space.

With \(\mathcal C_W(\mathbf r_1,\mathbf r_2)\equiv\langle W\mathcal H(\mathbf r_1),W\mathcal H(\mathbf r_2)\rangle_\Gamma\), the aperture-induced modification in the correlation kernel is
\begin{equation}
  \mathcal C_W(\mathbf r_1,\mathbf r_2)-
\mathcal C_\Gamma(\mathbf r_1,\mathbf r_2) =
\left\langle \mathcal H(\mathbf r_1),\left(W-I\right)\mathcal H(\mathbf r_2)\right\rangle_\Gamma
\label{eq:non_orthogonal}
\end{equation}
We denote this difference by \(\Delta\mathcal C_W\). It quantifies the aperture-induced modification of lateral and axial mode coupling relative to the complete-aperture reference. The resulting projection shadow is therefore deterministic: it arises from incomplete field access rather than from stochastic noise or computational error. Numerical suppression beyond what is supported by the measurements can induce physical overfitting by assigning unconstrained components within the null space~\cite{hansen2010discrete}. This may disrupt phase continuity and yield reconstructions inconsistent with the wave-propagation model. This projection shadow, therefore, provides a physical signature of observational incompleteness.

\begin{figure}[b]
    \centering
    \includegraphics[width=\columnwidth]{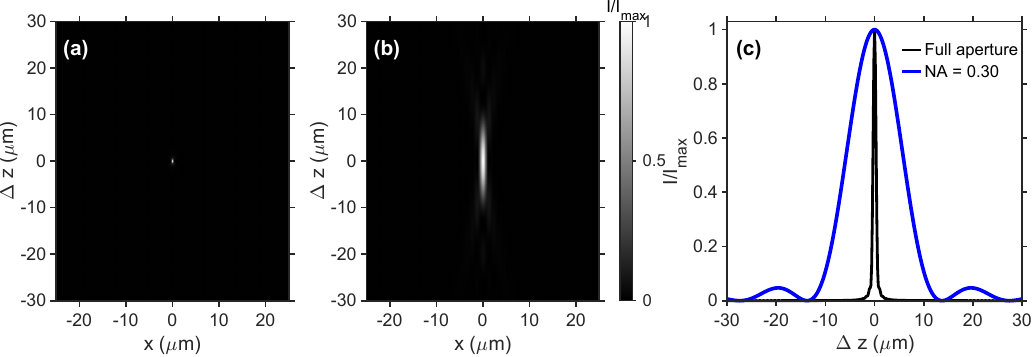}

    \vspace{-2mm}
    \caption{\textbf{Aperture-induced projection shadow in time-reversed wave evolution.} Independently normalized \(x\)-\(z\) intensities for
     (a) a complete-aperture reference and (b) finite-aperture observation (\(\mathrm{NA}=0.30\)). Finite angular acceptance yields the non-unitary effective observation map \(\mathcal H_A=W\mathcal H\) and broadens the axial point response into a deterministic projection shadow. (c) Corresponding normalized axial point-spread functions, with \(\Delta z=z-z_0\). Aperture truncation broadens the full width at half maximum from \(0.71~\mu\mathrm m\) to \(12.16~\mu\mathrm m\) (a \(17.1\)-fold increase).
}
    \label{fig:fig2}
\end{figure}

\begin{figure*}[t]
    \centering
    \includegraphics[width=0.99\textwidth]{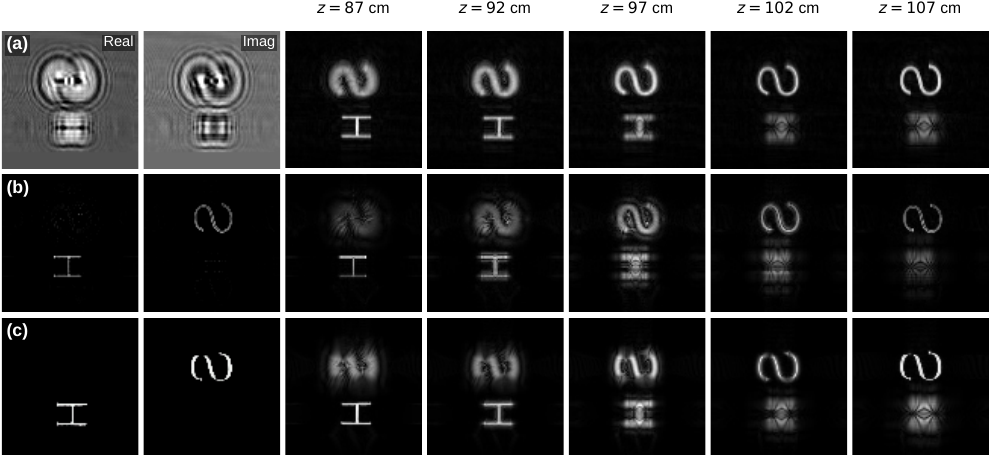}
    \vspace{-2mm}
    \caption{\textbf{Physical overfitting in algorithmic reconstructions.}
    (a) Experimental baseline $\mathcal{H}_A(z)^\dagger f$. The first two columns show the raw complex hologram $f$ (real and imaginary parts), followed by its direct back-propagation. (b),(c) Synthesized 3D wave fields $\mathcal{H}_A(z)^\dagger \mathcal{H}_A(z) u$ derived from the 2D focal estimates $u$ (first two columns) via regularized inverse imaging and U-Net, respectively. While enforcing spatial priors yields visually clean focal planes at target depths ($z = 87$ and 107 cm), it
    produces structured discrepancies at intermediate planes relative to the experimental adjoint reference.}
    \label{fig:fig3}
\end{figure*}

\section{Propagation-Resolved Physical Fidelity}

While the mathematical self-consistency of a wave field can be expressed by its departure from the projection manifold,
a rigorous physical criterion must be anchored to the experimental observation $f$ as well as the algorithmically reconstructed field $u$. Therefore, we define the operational TRS residual as:

\begin{equation}
    R_{\text{TRS}}(z) = \| \nabla \big[ \mathcal{H}_A(z)^\dagger \mathcal{H}_A(z)  u - \mathcal{H}_A(z)^\dagger f  \big] \|^2
\end{equation}
where $\mathcal H_A(z)=\mathcal W \mathcal H(z)$ is the finite-aperture observation operator at axial depth \(z\). Unlike conventional fidelity measures evaluated only on the 2D detection plane, \(R_{\mathrm{TRS}}(z)\) assesses the propagation-resolved 3D complex field. The measured-field reference, \(\mathcal H_A^\dagger(z)f\), retains the physical axial evolution of the aperture-induced projection shadow. In contrast, \(\mathcal H_A^\dagger(z)\mathcal H_A(z)u\) describes the corresponding forward--adjoint evolution of the algorithmically reconstructed field. Thus, \(R_{\mathrm{TRS}}(z)\) evaluates their discrepancy at each axial plane, providing a \(z\)-resolved test of physical fidelity. The projection shadow thereby serves as a physical reference, rather than as a residual to be suppressed. The gradient emphasizes spatially structured departures from the measured wave-propagation model, including deviations associated with unconstrained null-space components.

\section{Experimental Validation}

While the proposed theoretical framework and the $R_{\text{TRS}}(z)$ metric are universally applicable across coherent wave-based inverse problems, we select Optical Scanning Holography (OSH) as a representative experimental testbed. Among various coherent imaging modalities, OSH directly captures the twin-image-free complex field $f(x,y)$ of the object via heterodyne detection~\cite{Poon:09} (see Appendix A for the setup, implementation details, and data and code availability). This direct complex-field acquisition is critical, as it bypasses the non-linear errors inherent in iterative phase retrieval~\cite{Shechtman:15} and allows for the precise, linear implementation of the adjoint operator $\mathcal{H}_A(z)^\dagger$. In this architecture, the numerical aperture (NA) of the scanning objective deterministically restricts the spatial frequency bandwidth, perfectly instantiating the truncation operator $\mathcal{W}$. Consequently, OSH provides an ideal open-system environment with finite aperture observation to empirically validate the non-unitary projection shadow.

Defocus interference is commonly interpreted as inter-layer cross-talk from out-of-focus objects~\cite{Goodman:04}. To isolate the aperture-induced contribution, we simulate time-reversal back-propagation from a single point source [Fig.~\ref{fig:fig2}]. Even without adjacent scattering layers, finite angular acceptance broadens the axial point response. As the numerical aperture decreases, the truncated angular spectrum produces a deterministic, extended 3D PSF (see Appendix B for its evolution with NA and \(z\)). Thus, the defocus background does not require extrinsic inter-layer interference; it also arises as an aperture-induced projection shadow of a single-point response under finite-aperture observation.

Having isolated the aperture-induced contribution using a single-point simulation, we next examine whether the same physical constraint persists in an experimentally measured two-layer complex field. Using angular spectrum propagation, the adjoint reconstruction, \(\mathcal H_A^\dagger(z)f\), retains the aperture-limited axial structure across intermediate planes as shown in Fig.~\ref{fig:fig3}(a), providing the experimental reference.

The comparison is thus established between this reference and representative image-centric reconstructions obtained by regularized inverse imaging (a modified Tikhonov regularization)~\cite{Zhang:08} and a U-Net~\cite{ou:25}. It can be seen from the 2D focal estimates $u$ (first two columns of Figs.~\ref{fig:fig3}(b) and (c)) that both methods yield visually clean 2D reconstructions. However, propagating these estimates through the same aperture-limited forward--adjoint operator, ($\mathcal H_A^\dagger(z)\mathcal H_A(z)u$), reveals structured discrepancies at reconstruction planes relative to the experimental reference. These results show that suppressing the projection shadow can induce physical overfitting by assigning unconstrained null-space components, yielding reconstructions inconsistent with wave propagation.

\begin{figure}[t]
    \centering
    \includegraphics[width=0.95\columnwidth]{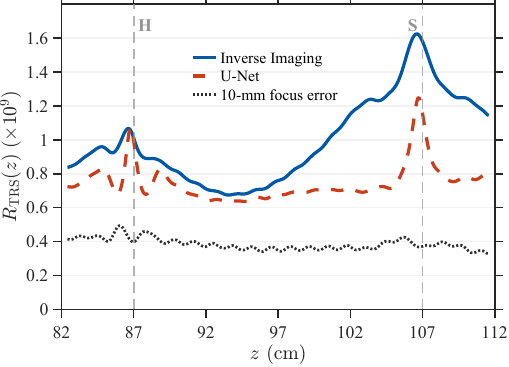}

    \vspace{-2mm}
    \caption{\textbf{Quantitative assessment of physical fidelity.}
    The TRS residual ($R_{\mathrm{TRS}}(z)$) is evaluated along the propagation axis. Direct back-propagation with a deliberate 10-mm defocus error maintains a low residual, indicating consistency with the forward--adjoint wave-propagation model. In contrast, inverse imaging (modified Tikhonov) reconstruction and the U-Net exhibit pronounced residual peaks near the estimated focal planes of H and S, where they yield their clearest 2D reconstructions. The results reveal 3D complex-field discrepancies that are not captured by two-dimensional image quality alone.}
    \label{fig:fig4}
\end{figure}

To quantify these discrepancies, we evaluate the physical fidelity of the reconstructed wave fields using the TRS residual metric, \(R_{\mathrm{TRS}}(z)\) [Fig.~\ref{fig:fig4}]. As a physical control, direct back-propagation with a deliberate \(10\)-mm defocus error maintains a low residual, showing that conventional focal inaccuracies can remain consistent with the forward--adjoint wave-propagation model. In contrast, the fields synthesized by inverse imaging and the U-Net exhibit pronounced residual peaks near the estimated focal planes of H and S, where the algorithms produce their cleanest 2D reconstructions. This behavior persists in ideal operator simulations and is distinct from conventional defocus errors (see Appendices C and D). By embedding the measured axial evolution of the projection shadow in the reference field, \(R_{\mathrm{TRS}}(z)\) turns this aperture-induced residual into a \(z\)-resolved benchmark of physical fidelity. These results show that detector-plane optimization can conceal discrepancies in the propagation-resolved 3D complex field. By evaluating the measured and reconstructed field evolutions across \(z\), \(R_{\mathrm{TRS}}(z)\) provides an operational metric of physical fidelity beyond 2D image quality.

\section{Conclusion}

In conclusion, we show that the defocus background under finite-aperture observation in open systems is not merely a stochastic artifact, but a deterministic projection shadow reflecting irretrievable information loss and its associated physical uncertainty. Suppressing this residual beyond what is supported by the measurements can yield visually clean reconstructions that are inconsistent with wave propagation. Thus, 2D image metrics alone may favor solutions that lack physical fidelity. The \(R_{\mathrm{TRS}}(z)\) metric provides an operator-based, object-space test of physical fidelity by comparing the forward--adjoint reconstruction of the algorithmic field with the adjoint reconstruction determined by the measured complex field. It identifies structured discrepancies associated with aperture-limited TRS breaking and unconstrained null-space components.

Rather than treating the aperture-induced projection shadow as a nuisance to be removed, $R_{\mathrm{TRS}}(z)$ uses its measured axial evolution as a physical reference. This metric thus extends physical-fidelity assessment from detector-plane agreement to propagation-resolved complex-field consistency. More broadly, these results motivate computational architectures for wave-based inverse problems that complement statistical priors with operator-based consistency under wave propagation.

\section*{Acknowledgments}

This work was supported by Sichuan Science and Technology Program (No.~2026NSFSC0398) and Guangdong Basic and Applied Basic Research Foundation for Meteorological Joint Funds (No.~2024A1515510017).

\appendix
\setcounter{figure}{4}
\section*{Appendix A: Experimental Setup and Algorithm Implementation}

To rigorously capture the complex wave field and completely circumvent the nonlinear artifacts inherent to traditional phase retrieval algorithms, our experiments utilize heterodyne optical scanning holography (OSH). Unlike intensity-only imaging schemes, this heterodyne detection strategy deterministically records both amplitude and phase, providing a pure complex-valued hologram as the physical baseline for our $R_{\text{TRS}}(z)$ metric.

\begin{figure}[b]
  \centering
  \includegraphics[width=0.98\columnwidth]{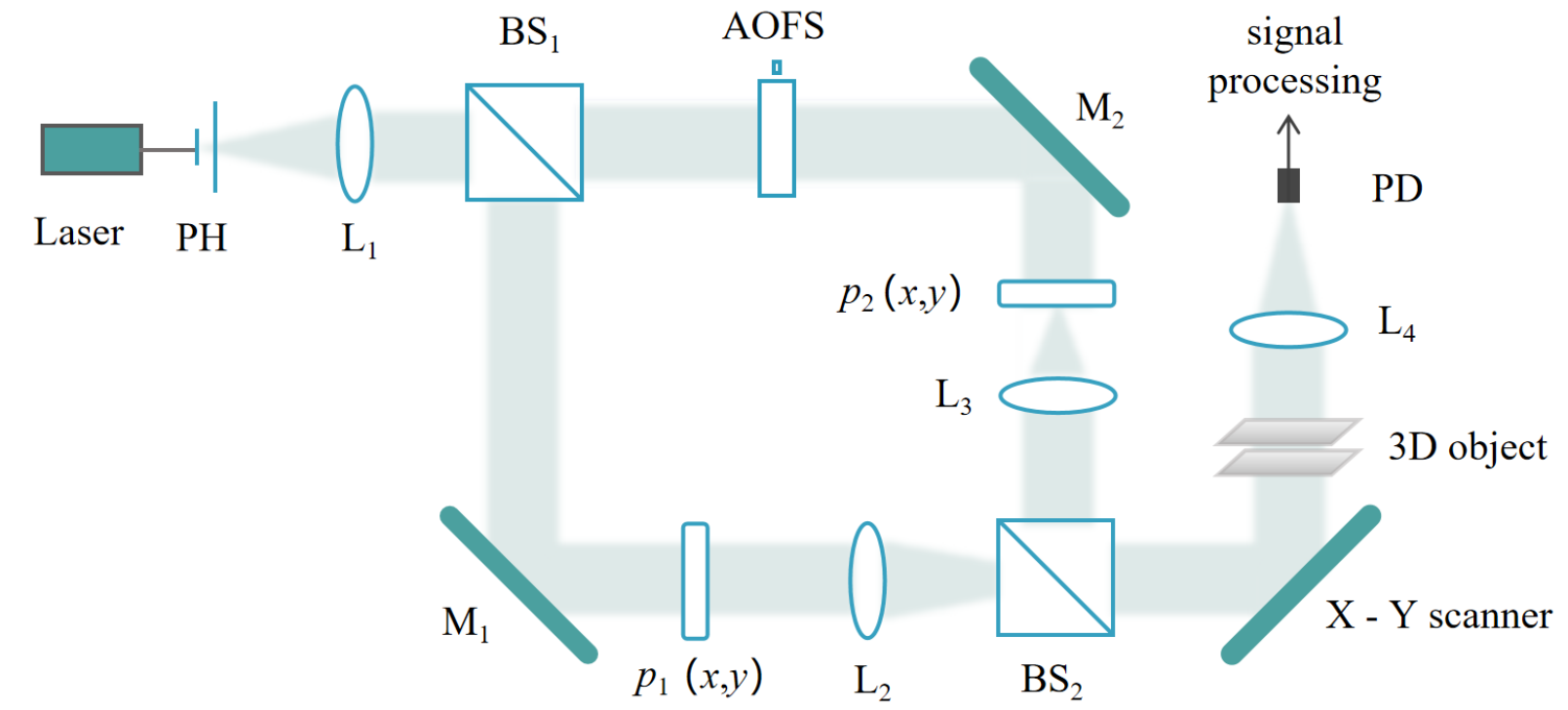}
  \caption{Experimental configuration of the heterodyne optical scanning holography system. PH: pinhole; BS: beam splitter; M: mirror; AOFS: acousto-optic frequency shifter; L: lens; PD: photodetector; $p(x,y)$: pupil.}
  \label{fig:app-setup}
\end{figure}

The optical configuration is illustrated in Fig.~\ref{fig:app-setup}. A He-Ne laser with a central wavelength of 632.8 nm is spatially filtered by a pinhole (PH) and collimated by lens L1, establishing a collimated beam diameter of $D = 25$ mm before being divided into two interferometric arms by the first beam splitter (BS1). In the lower arm, the beam is reflected by mirror M1 and passes through the uniform pupil $p_1(x,y) = 1$ and lens L2. In the upper arm, the beam undergoes a temporal frequency shift through an acousto-optic frequency shifter (AOFS), is reflected by mirror M2, and then passes through the pinhole pupil $p_2(x,y) = \delta(x,y)$ and lens L3. Here, lenses L2 and L3 govern the optical scanning formulation, both possessing a focal length of $f = 500$ mm. The two beams are recombined at BS2 to generate the spatially modulated interference pattern, which is scanned across the three-dimensional target by an X-Y scanner. The target consists of two transparent slides located at $z_1 = 87$ cm and $z_2 = 107$ cm, respectively. The transmitted optical signal passes through lens L4 and is detected by the single-pixel photodetector (PD), followed by standard heterodyne signal processing to directly recover the complex hologram.

\vspace{2mm}

Mathematically, for a three-dimensional target discretized into $N$ longitudinal planes, the acquired complex hologram $g(x,y)$ is formulated as a linear superposition:

\begin{equation}
    g(x,y) \approx \sum_{i=1}^{N}
    \left(
    \vert{}O(x,y; z_i)\vert{}^2 \otimes h(x,y; z_i)
    \right)
\end{equation}
where $O(x,y; z_i)$ denotes the complex amplitude of the $i$-th depth layer, $h(x,y; z_i)$ represents the depth-dependent spatial impulse response, and $\otimes$ is the 2D convolution operator. Specifically, utilizing the wavenumber $k = 2\pi/\lambda$, this spatial impulse response is expressed as~\cite{Poon:09}:

\begin{equation}
 h(x,y; z_i) = \frac{j k}{2 \pi z_i} \exp\left[ -j \frac{k}{2 z_i} (x^2 + y^2) \right]
\end{equation}
Crucially, in the physical system, this ideal free-space propagation is strictly band-limited by the finite collimated beam diameter $D = 25$ mm. This physical aperture acts as the spatial truncation operator $\mathcal{W}$ discussed in the main text, mathematically restricting the spatial coordinates $(x,y)$ from an infinite plane to a finite bounded domain. It is precisely this finite spatial support that strictly prevents the complete retrieval of the complex wave field, inherently breaking the time-reversal symmetry of the wave evolution.

By numerically reconstructing this hologram at a specific focal plane $z_l$ through cross-correlation with the conjugate spatial impulse response, the output intensity field $I_{\text{out}}(x,y; z_l)$ becomes:
\begin{equation}
\begin{aligned}
I_{\text{out}}(x,y; z_l)
={}& \vert O(x,y; z_l)\vert^2 \otimes h(x,y; z_l) \\
&\quad \otimes h^*(x,y; z_l) \\
&+ \sum_{i \neq l} \vert O(x,y; z_i)\vert^2 \otimes h(x,y; z_i) \\
&\quad \otimes h^*(x,y; z_l)
\end{aligned}
\end{equation}
where the superscript $*$ indicates complex conjugation. Governed by the aforementioned spatial truncation $\mathcal{W}$, the entire reconstructed field is fundamentally band-limited. Consequently, the first term retrieves the in-focus structural information subject to a diffraction-limited physical resolution. Concurrently, the second summation explicitly formulates the defocus interference originating from all out-of-focus layers, the precise projection shadow analyzed in the main text.

\bigskip

\section*{Appendix B: Analytical 3D Point Spread Function Under Finite Apertures}

\begin{figure}[t]
  \centering
  \includegraphics[width=0.98\columnwidth]{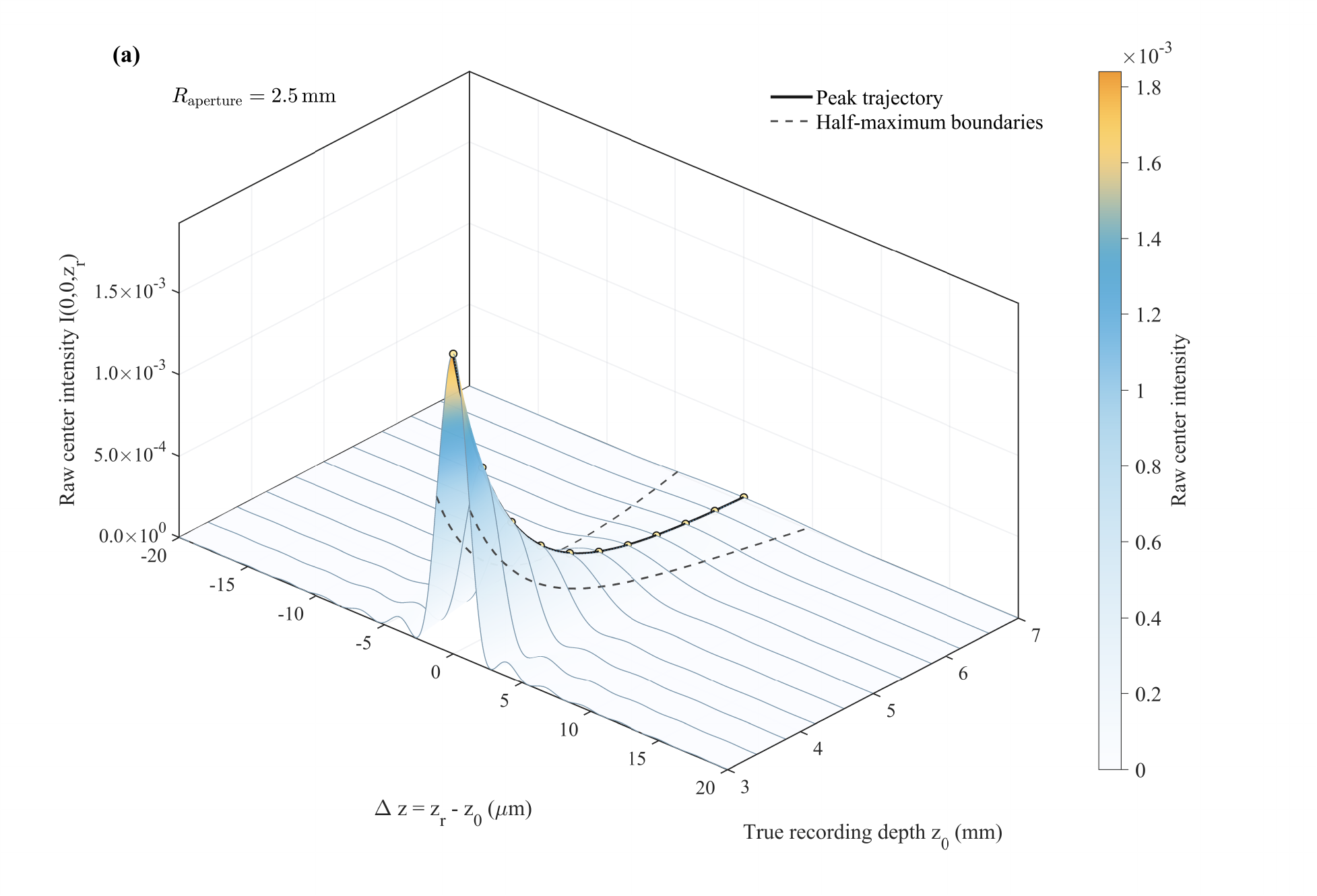}
  \includegraphics[width=0.98\columnwidth]{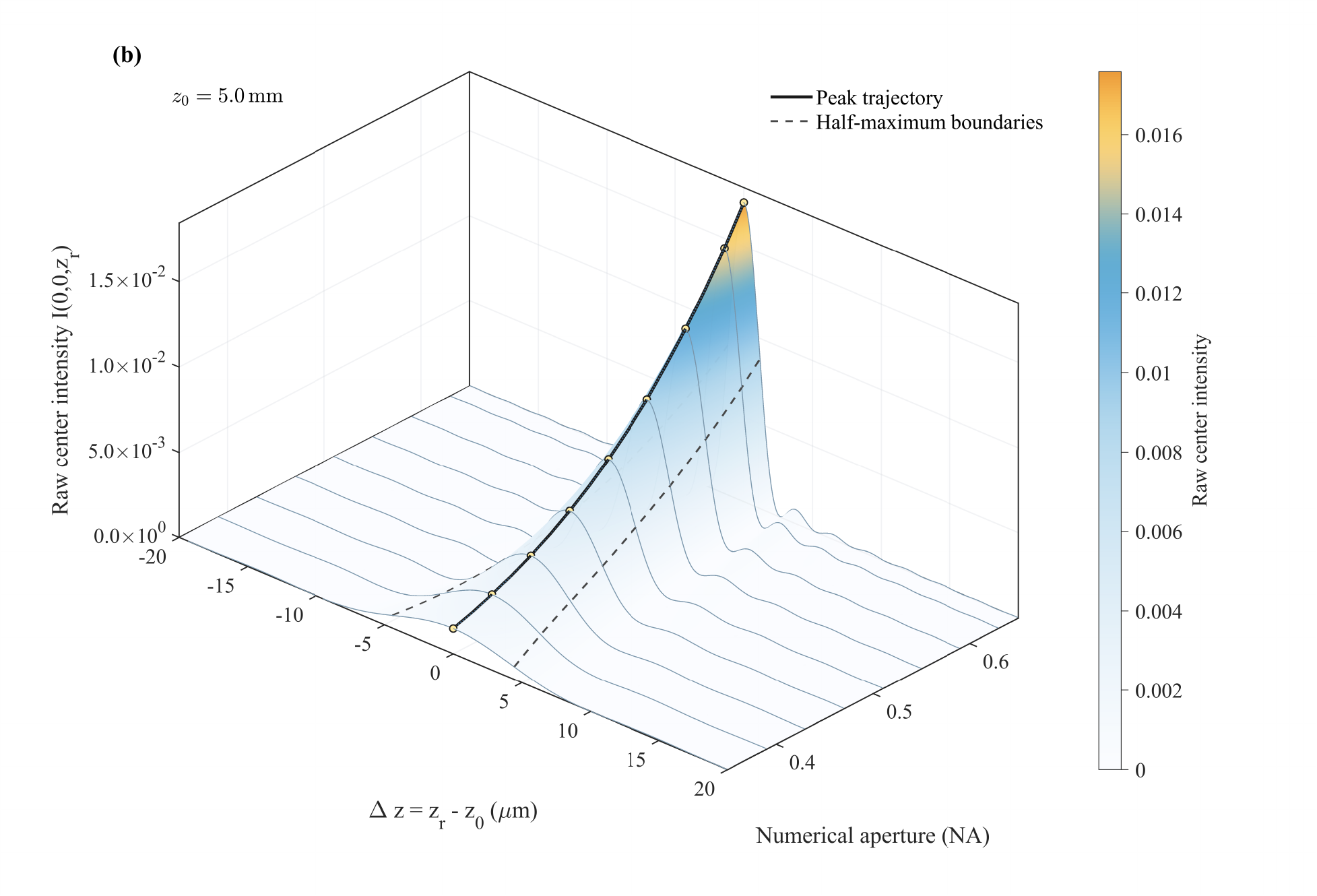}
  \caption{Axial point-source responses under finite-aperture constraints. (a) Fixed physical aperture $R_{\text{aperture}} = 2.5$~mm with varying recording depth $z_0$. (b) Fixed $z_0 = 5.0$~mm with varying numerical aperture (NA). Solid curves indicate peak trajectories, and dashed curves indicate half-maximum boundaries.}
  \label{fig:app-psf}
\end{figure}

To formally define the projection shadow induced by finite apertures, we derive the 3D point spread function of a point source based on scalar diffraction theory~\cite{Goodman:04}. Consider a coherent point source with wavelength $\lambda$. In an open wave system, the finite numerical aperture (NA) of the observation plane acts as a spatial frequency truncation operator, imposing a hard physical cutoff at the maximum spatial frequency $f_{r,\text{max}} = \text{NA}/\lambda$. Under the paraxial approximation, the complex amplitude $U(0,0,\Delta z)$ along the optical axis at a defocus distance $\Delta z = z_r - z_0$ is described by the following integral:

\begin{equation}
    U(0,0,\Delta z) = C \int_0^{\text{NA}/\lambda} \exp\left[ik\Delta z \left(1 - \frac{\lambda^2 f_r^2}{2}\right)\right] f_r df_r
\end{equation}
Here, $k = 2\pi/\lambda$ is the wave number, $f_r$ is the radial spatial frequency, and $C$ is a constant determined by system transmittance. Evaluating this integral and calculating its squared modulus yields the axial intensity distribution~\cite{Born_principle_of_optics_1999}:

\begin{equation}
    I(0,0,\Delta z) = I_0 \operatorname{sinc}^2 \left( \frac{\pi \text{NA}^2 \Delta z}{2 \lambda} \right)
\end{equation}
where $I_0$ represents the peak intensity at the focal plane $\Delta z = 0$. By adopting the standard sinc function convention $\text{sinc}(x) = \sin(x)/x$, this analytical formulation demonstrates that defocus interference is a deterministic coherent energy spread resulting from high-frequency information loss rather than a stochastic disturbance. The full width at half maximum (FWHM) boundaries follow an inverse square relationship with NA:

\begin{equation}
    \Delta z_{\text{FWHM}} \approx \frac{1.772 \lambda}{\text{NA}^2}
\end{equation}
As NA decreases, spatial bandwidth is truncated and the axial response deterministically broadens. Fig.~\ref{fig:app-psf} visualizes this analytical evolution via two complementary numerical angular-spectrum simulations. Panel (a) shows the axial response for a fixed physical aperture $R_{\text{aperture}} = 2.5$ mm while varying the true recording depth $z_0$. Here, the effective numerical aperture varies according to $\text{NA}_{\text{eff}} = R_{\text{aperture}} / \sqrt{R_{\text{aperture}}^2 + z_0^2}$. The simulation also retains the fundamental $1/z_0$ field-amplitude attenuation, so the variation of the raw peak intensity reflects both finite-aperture truncation and propagation-distance attenuation. Conversely, panel (b) shows the response at a fixed recording depth $z_0 = 5.0$ mm while varying the numerical aperture. In both panels, the solid black curve denotes the peak trajectory, and the dashed curves delineate the half-maximum boundaries. The high-frequency ripples extending along the base represent the coherent diffraction features associated with finite-aperture spatial-frequency truncation, which manifest as the projection shadow in the wave-based inverse problem.

This aperture-induced structure defines the measurement-supported axial response. Reconstruction methods that enforce sparsity, smoothness, or learned image priors can suppress part of this structure in selected two-dimensional planes. When such estimates are propagated through the same finite-aperture operator, however, they can deviate from the measured propagation-resolved complex field through unconstrained null-space components. This behavior motivates the \(R_{\mathrm{TRS}}(z)\) metric introduced in the main text.

\vspace{2mm}

\section*{Appendix C: Discriminating Physical Defocus from Propagation-Resolved Inconsistency}

\begin{figure*}[t]
  \centering
  \includegraphics[width=0.62\textwidth]{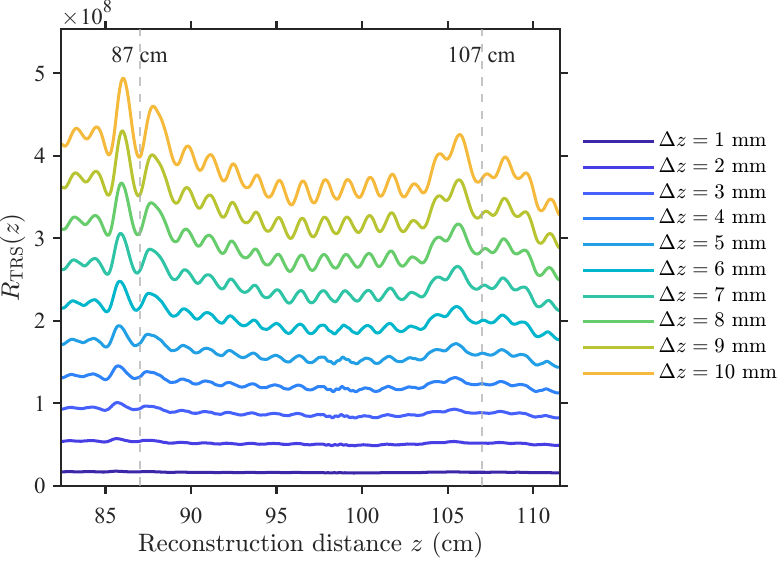}
  \caption{Response of \(R_{\mathrm{TRS}}(z)\) to controlled axial defocus. Curves show the residual for reference reconstructions with axial offsets \(\Delta z=1\)--\(10~\mathrm{mm}\). Increasing defocus elevates the overall residual and produces smooth, propagation-extended oscillations. This continuous defocus response differs from the localized residual peaks near the estimated focal planes produced by inverse imaging and the U-Net in Fig.~\ref{fig:fig4}, supporting the specificity of            \(R_{\mathrm{TRS}}(z)\) to propagation-resolved complex-field discrepancies.}
  \label{fig:app-defocus}
\end{figure*}

To determine whether \(R_{\mathrm{TRS}}(z)\) distinguishes ordinary physical defocus from algorithm-induced discrepancies, we impose controlled axial offsets \(\Delta z=1\)--\(10~\mathrm{mm}\) on the reference reconstruction. These offsets model uncertainty in the estimated axial position of a target layer. Figure~\ref{fig:app-defocus} shows that increasing physical defocus raises the overall residual and produces smooth, spatially extended oscillations along the propagation axis. This behavior follows from the continuous evolution of the aperture-limited wave field under defocus and is qualitatively distinct from the localized residual peaks observed near the focal planes of the inverse-imaging and U-Net reconstructions in Fig.~\ref{fig:fig4}.

In practical holographic reconstruction, axial localization is typically more accurate than the deliberately conservative \(1\)--\(10~\mathrm{mm}\) offsets considered here~\cite{wu2018extended,ou:26}. The control therefore demonstrates that the pronounced peaks of \(R_{\mathrm{TRS}}(z)\) in Fig.~\ref{fig:fig4} cannot be attributed solely to conventional defocus. Instead, they indicate discrepancies between the propagation of the reconstructed complex field and the evolution supported by the measured field, consistent with contributions from unconstrained null-space components. This comparison establishes \(R_{\mathrm{TRS}}(z)\) as a metric that distinguishes propagation-resolved physical inconsistency from ordinary focal mislocalization.

\section*{Appendix D: Algorithmic Reconstructions under Idealized Finite-Aperture Simulations}

To determine whether the observed discrepancies depend on experimental noise, target-specific structure, or operator mismatch, we construct an idealized dual-layer target with layer A at \(z=87~\mathrm{cm}\) and layer B at \(z=107~\mathrm{cm}\). Its complex hologram is generated using the same wavelength, aperture, sampling, and propagation model as in the experiment. The identical finite-aperture operator \(\mathcal H_A\) is used to generate the data, reconstruct the fields, and evaluate \(R_{\mathrm{TRS}}(z)\). Thus, while the incomplete field access imposed by the finite aperture is retained, measurement noise, additional aberrations, and mismatch between the physical and modeled observation operators are excluded.

\begin{figure*}[p]
  \centering
  \includegraphics[width=0.82\textwidth]{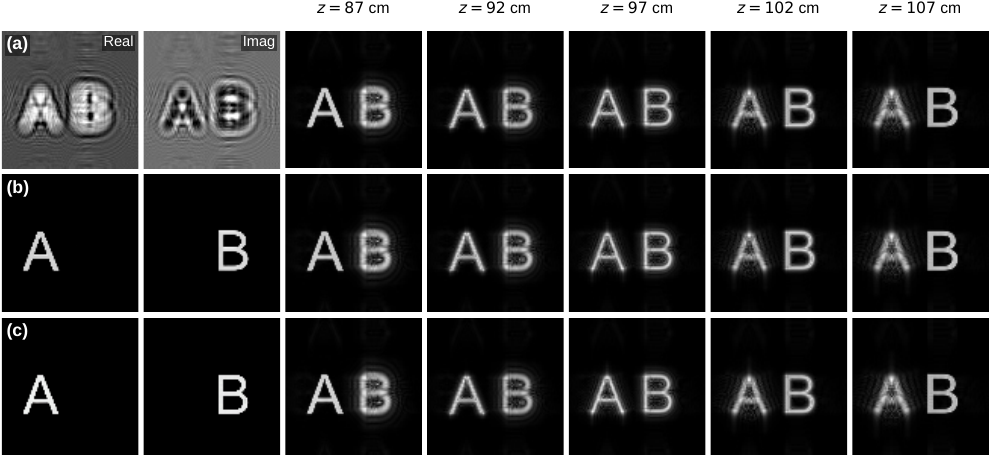}
  \caption{Idealized finite-aperture simulation of a dual-layer target. (a) Simulated complex hologram and adjoint reference field, \(\mathcal H_A^\dagger(z)f\), shown over the indicated reconstruction planes. Layer A and layer B are located at \(z=87~\mathrm{cm}\) and \(z=107~\mathrm{cm}\), respectively. (b,c) Forward--adjoint fields, \(\mathcal H_A^\dagger(z)\mathcal H_A(z)u\), obtained from U-Net and inverse imaging reconstructions, respectively. All three reconstructions remain visually similar across the displayed planes. Their propagation-resolved differences are further quantified by \(R_{\mathrm{TRS}}(z)\), demonstrating that the effect persists without experimental noise or operator mismatch.}
  \label{fig:app-simulation}
  \vspace{2mm}
  \includegraphics[width=0.82\textwidth]{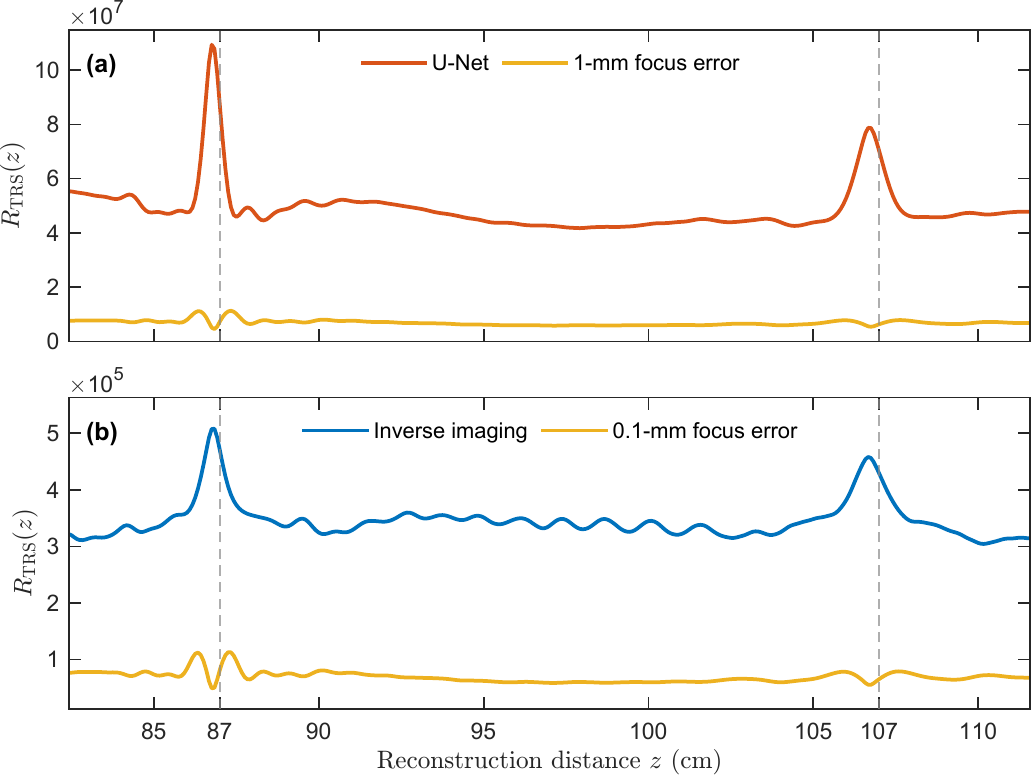}
  \caption{\(R_{\mathrm{TRS}}(z)\) under idealized finite-aperture simulations. (a) The U-Net reconstruction is compared with a 1-mm physical-defocus reference. (b) Inverse imaging reconstruction is compared with a 0.1-mm physical-defocus reference, selected to resolve its smaller residual scale. The different defocus offsets are used for within-method scale comparison. In both panels, physical defocus yields a smooth propagation-extended residual, whereas the algorithmic reconstructions exhibit localized peaks near the estimated target planes.}
  \label{fig:app-residual}
\end{figure*}

Fig.~\ref{fig:app-simulation}(a) shows the simulated complex hologram and the adjoint reference field, \(\mathcal H_A^\dagger(z)f\). Figures~\ref{fig:app-simulation}(b) and~\ref{fig:app-simulation}(c) show the corresponding forward--adjoint fields, \(\mathcal H_A^\dagger(z)\mathcal H_A(z)u\), obtained from the U-Net and inverse-imaging (modified-Tikhonov) reconstructions, respectively. In contrast to the experimental results, the three reconstructions in the idealized simulation remain visually similar across the displayed axial planes. Visual inspection of the reconstructed intensities alone therefore provides little basis for distinguishing their physical fidelity. The propagation-resolved comparison and \(R_{\mathrm{TRS}}(z)\), however, reveal differences between the algorithmic fields and the simulated reference evolution that are not apparent from the two-dimensional images.

Fig.~\ref{fig:app-residual} quantifies these discrepancies using \(R_{\mathrm{TRS}}(z)\). For the U-Net, we compare the reconstructed field with a 1-mm physical-defocus reference, whose residual magnitude resolves the relevant scale of the U-Net response. For inverse imaging reconstruction, we use a 0.1-mm physical-defocus reference to resolve its  smaller residual scale. These offsets are selected for within-method comparison and are not intended as a quantitative calibration between the two panels. In both cases, physical defocus produces a smooth propagation-extended residual, whereas the reconstructed fields exhibit localized residual peaks near the estimated target planes. Because these peaks persist without experimental noise or operator mismatch, they indicate discrepancies between the propagation of the algorithmic reconstructions and that supported by the finite-aperture reference field. This result supports the interpretation of physical overfitting as the assignment of unconstrained components in the null space when image-centric reconstruction suppresses the projection shadow.

\section*{Data and Code Availability}
The data and code supporting the U-net network from our former publication are publicly available at \url{https://github.com/CrackCzi/Defocus-and-speckle-noise-suppression-in-optical-scanning-holography}

\clearpage

\bibliography{arxiv_references}

@article{Fink1997Time,
  title={Time Reversed Acoustics},
  author={Fink, Mathias},
  journal={Reports on Progress in Physics},
  volume={50},
  number={3},
  pages={34-40},
  year={1997},
}

@book{Born_principle_of_optics_1999, 
place={Cambridge}, 
edition={7}, 
title={Principles of Optics: Electromagnetic Theory of Propagation, Interference and Diffraction of Light}, publisher={Cambridge University Press}, 
author={Born, Max and Wolf, Emil}, 
year={1999}
}

@book{bertero1998introduction,
  author    = {Bertero, Mario and Boccacci, Patrizia},
  title     = {Introduction to Inverse Problems in Imaging},
  publisher = {CRC Press},
  year      = {1998},
  doi       = {10.1201/9780367806941}
}

@article{RUDIN:1992,
title = {Nonlinear total variation based noise removal algorithms},
journal = {Physica D: Nonlinear Phenomena},
volume = {60},
number = {1},
pages = {259-268},
year = {1992},
issn = {0167-2789},
doi = {https://doi.org/10.1016/0167-2789(92)90242-F},
url = {https://www.sciencedirect.com/science/article/pii/016727899290242F},
author = {Leonid I. Rudin and Stanley Osher and Emad Fatemi},
}

@ARTICLE{Candes:06,
author={Candes, E.J. and Romberg, J. and Tao, T.},
journal={IEEE Trans. Inf. Theory},
title={Robust uncertainty principles: exact signal reconstruction from highly incomplete frequency information},
volume={52},number={2},pages={489--509},
year=2006}

@article{Barbastathis:19,
author = {George Barbastathis and Aydogan Ozcan and Guohai Situ},
journal = {Optica},
number = {8},
pages = {921--943},
publisher = {Optica Publishing Group},
title = {On the use of deep learning for computational imaging},
volume = {6},
month = {Aug},
year = {2019},
url = {https://opg.optica.org/optica/abstract.cfm?URI=optica-6-8-921},
doi = {10.1364/OPTICA.6.000921},
}

@article{Zeng:21,
author = {Tianjiao Zeng and Yanmin Zhu and Edmund Y. Lam},
journal = {Opt. Express},
number = {24},
pages = {40572--40593},
publisher = {Optica Publishing Group},
title = {Deep learning for digital holography: a review},
volume = {29},
month = {Nov},
year = {2021},
url = {https://opg.optica.org/oe/abstract.cfm?URI=oe-29-24-40572},
doi = {10.1364/OE.443367},
}

@article{Carminati:07,
author = {R. Carminati and R. Pierrat and J. de Rosny and M. Fink},
journal = {Opt. Lett.},
number = {21},
pages = {3107--3109},
publisher = {Optica Publishing Group},
title = {Theory of the time reversal cavity for electromagnetic fields},
volume = {32},
month = {Nov},
year = {2007},
url = {https://opg.optica.org/ol/abstract.cfm?URI=ol-32-21-3107},
doi = {10.1364/OL.32.003107},
}

@ARTICLE{Fink:92,
  author = {Fink, M.},
  title = {Time reversal of ultrasonic fields. I. Basic principles},
  journal = {IEEE T. Ultrason. Ferr.},
  year = {1992},
  volume = {39},
  pages = {555--566},
  number = {5}
}

@ARTICLE{Fink:93,
  author = {Fink, M.},
  title = {Time-reversal mirrors},
  journal = {J. Phys. D: Appl. Phys.},
  year = {1993},
  volume = {26},
  pages = {1333--1350},
  number = {9}
}

@book{hansen2010discrete,
  author    = {Hansen, P. C.},
  title     = {Discrete Inverse Problems: Insight and Algorithms},
  publisher = {SIAM},
  address   = {Philadelphia},
  year      = {2010},
  doi       = {10.1137/1.9780898718836},
  series    = {Fundamentals of Algorithms}
}

@ARTICLE{Poon:09,
  author = {Ting-Chung Poon},
  title = {Optical Scanning Holography - A Review of Recent Progress},
  journal = {Journal of the Optical Society of Korea},
  year = {2009},
  volume = {13},
  pages = {406--415},
  number = {4}
}

@ARTICLE{Shechtman:15,
  author={Shechtman, Yoav and Eldar, Yonina C. and Cohen, Oren and Chapman, Henry Nicholas and Miao, Jianwei and Segev, Mordechai},
  journal={IEEE Signal Processing Magazine}, 
  title={Phase Retrieval with Application to Optical Imaging: A contemporary overview}, 
  year={2015},
  volume={32},
  number={3},
  pages={87-109},
  doi={10.1109/MSP.2014.2352673}
  }

@BOOK{Goodman:04,
  title = {Introduction to Fourier Optics},
  publisher = {Roberts and Company},
  year = {2004},
  author = {Joseph W. Goodman},
  edition = {Third}
}

@ARTICLE{Zhang:08,
  author = {Xin Zhang and Edmund Y. Lam and Ting-Chung Poon},
  title = {Reconstruction of Sectional Images in Holography Using Inverse Imaging},
  journal = {Opt. Express},
  year = {2008},
  volume = {16},
  pages = {17215--17226},
  number = {22},
}

@article{ou:25,
title = {Defocus and speckle noise suppression in optical scanning holography},
journal = {Optics \& Laser Technology},
volume = {190},
pages = {113246},
year = {2025},
issn = {0030-3992},
doi = {https://doi.org/10.1016/j.optlastec.2025.113246},
url = {https://www.sciencedirect.com/science/article/pii/S0030399225008370},
author = {Ruiwei Xu and Haiyan Ou and Edmund Y. Lam},

}

@article{wu2018extended,
  title={Extended depth-of-field in holographic imaging using deep-learning-based autofocusing and phase recovery},
  author={Wu, Yichen and Rivenson, Yair and Zhang, Yibo and Wei, Zhensong and G{\"u}naydin, Harun and Lin, Xing and Ozcan, Aydogan},
  journal={Optica},
  volume={5},
  number={6},
  pages={704--710},
  year={2018},
  publisher={Optica Publishing Group}
}

@article{ou:26,
author = {Ye Liu and Bing-Zhong Wang and Edmund Lam and Haiyan Ou},
journal = {Opt. Lett.},
number = {10},
pages = {2736--2739},
publisher = {Optica Publishing Group},
title = {Autofocus in digital holography with time reversal and depth-dependent sampling},
volume = {51},
month = {May},
year = {2026},
url = {https://opg.optica.org/ol/abstract.cfm?URI=ol-51-10-2736},
doi = {10.1364/OL.591504},
}

\end{document}